\documentclass[conference]{IEEEtran}
\IEEEoverridecommandlockouts
\usepackage{soul}
\usepackage{cite}
\usepackage{amsmath,amssymb,amsfonts,mathtools}
\usepackage{algorithmic}
\usepackage{graphicx}
\usepackage{textcomp}
\usepackage{xcolor}
\usepackage{multirow}
\usepackage{subfigure}
\usepackage[symbol]{footmisc}
\def\BibTeX{{\rm B\kern-.05em{\sc i\kern-.025em b}\kern-.08em
    T\kern-.1667em\lower.7ex\hbox{E}\kern-.125emX}}
\begin{document}

\title{Robust Attitude Estimation with Quaternion Left-Invariant EKF and Noise Covariance Tuning\\

}

\author{\IEEEauthorblockN{Yash Pandey\IEEEauthorrefmark{1}}
\IEEEauthorblockA{\textit{Department of Electronics Engineering} \\
\textit{Harcourt Butler Technical University}\\
Kanpur, India \\
yashpanknp@gmail.com}

\and
\IEEEauthorblockN{Rahul Bhattacharyya\IEEEauthorrefmark{1}}
\IEEEauthorblockA{\textit{Department of Electrical Engineering} \\
\textit{Indian Institute of Technology, Kanpur} \\
Kanpur, India \\
rahulbhatta0@gmail.com}

\and
\IEEEauthorblockN{Yatindra Nath Singh}
\IEEEauthorblockA{\textit{Department of Electrical Engineering} \\
\textit{Indian Institute of Technology, Kanpur} \\
Kanpur, India \\
ynsingh@iitk.ac.in}
}
\maketitle
\footnotetext[1]{Contributed Equally}

\begin{abstract}
Accurate estimation of noise parameters is critical for optimal filter performance, especially in systems where true noise parameter values are unknown or time-varying. This article presents a quaternion left-invariant extended Kalman filter (LI-EKF) for attitude estimation, integrated with an adaptive noise covariance estimation algorithm. By employing an iterative expectation-maximization (EM) approach, the filter can effectively estimate both process and measurement noise covariances. Extensive simulations demonstrate the superiority of the proposed method in terms of attitude estimation accuracy and robustness to initial parameter misspecification. The adaptive LI-EKF’s ability to adapt to time-varying noise characteristics makes it a promising solution for various applications requiring reliable attitude estimation, such as aerospace, robotics, and autonomous systems.
\end{abstract}

\begin{IEEEkeywords}
State Estimation, Kalman Filtering, Invariant Theory, Maximum Likelihood Estimation, Adaptive Filtering, Non-linear Filtering
\end{IEEEkeywords}

\section{Introduction}
The Kalman filter and its variants have been widely used in various estimation and control applications, particularly in the field of aerospace engineering \cite{schmidt1981,oh2006}, navigation systems \cite{grewal2020} and robotics \cite{chen2011,jetto1999,moore2016}. Rudolf E. Kalman\cite{kalman1960new} demonstrated that his proposed filter provides an optimal state estimate by minimizing the mean squared error (MSE) when the system is \textit{linear} and the noise is \textit{Gaussian}.
This filter, which was later named after him, is commonly referred to as the additive Kalman filter. The additive Kalman filter was originally developed for linear systems. However, when dealing with nonlinear systems, such as attitude estimation, the extended Kalman filter (EKF) was introduced to linearize the system around the current state estimate. The multiplicative EKF (MEKF) emerged as a solution to handle attitude estimation more effectively, particularly when using quaternions to represent orientation. The MEKF operates on the error between the true and estimated quaternions, addressing issues related to over-parameterization and singularities \cite{markley2014fundamentals}. Building upon these advancements, the left-invariant EKF (LI-EKF) has gained attention due to its ability to preserve the geometric structure of the state space, leading to improved consistency and better convergence properties \cite{brossard2018invariant}.\par Accurate specification of process noise covariance ($Q$) and measurement noise covariance ($R$) matrices is crucial for Kalman filter performance. These matrices represent uncertainties in the system model and sensor measurements. $Q$ and $R$ are often unknown or time-varying, potentially causing sub-optimal performance or filter divergence if incorrectly specified. This has prompted research in adaptive filtering techniques to estimate or adjust $Q$ and $R$ online during operation \cite{mehra1970,mohamed1999,shumway_em,bavdekar_em,ananthasayanam2016}. The accurate estimation of these noise parameters is particularly crucial in attitude estimation problems, where system non-linearities and measurement complexities can exacerbate the effects of incorrectly estimated noise covariances. The LI-EKF is particularly well-suited for attitude estimation problems, where the state evolves on the special orthogonal group SO(3) or its double cover, the unit quaternions \cite{wu2016fast}. In this article, we focus on the estimation of $Q$ and $R$ within the LI-EKF framework for attitude estimation. We use a quaternion-based state dynamics model to estimate attitude. We focus on the stability and convergence of noise covariance estimates, the accuracy of attitude estimates, and the sensitivity to initial parameter values.
By analyzing these factors, we aim to provide a comprehensive understanding of the LI-EKF’s behavior in attitude estimation and the impact of adaptive noise covariance estimation on its performance. The main contribution of this article is as follows:
\begin{itemize}
\item This article proposes a robust LI-EKF for attitude estimation, enhanced by an iterative expectation-maximization (EM) algorithm. The EM algorithm is employed to estimate both the $Q$ and $R$, allowing them to adapt to the true noise characteristics of the system.
\end{itemize}
\section{Background and Problem Formulation}
\subsection{Kalman Filtering}
The extended Kalman filter (EKF) is a modification of the popular Kalman filter that provides a sub-optimal solution to non-linear estimation problems by linearizing the model equations around the present estimate \cite{simon_ekf}. Consider the system
\begin{align}
    \label{ekf_prop}
    x_{k+1} &= f(x_k, w_k), \\
    \label{ekf_meas}
    z_k &= h(x_k, v_k).
\end{align}
$x_k$ is the state-vector and $z_k$ is the measurement-vector at time step $k$. $w_k$ and $v_k$ are model noise and measurement noise vectors defined by 
\begin{align}
    \label{filter_noise}
    w_k\sim\mathcal{N}(0, Q), \quad v_k\sim\mathcal{N}(0, R).
\end{align}
$Q$ and $R$ are corresponding noise covariance matrices for $w_k$ and $v_k$  respectively.

The system is linearized by computing the Jacobian of the functions $f$ and $h$ about the present estimate i.e.,
\begin{align}
    F_k &= \frac{\partial f}{\partial x}\Big|_{x = \hat x_k^+}, & G_k &= \frac{\partial f}{\partial w}\Big|_{x = \hat x_k^+},\\
    H_k &= \frac{\partial h}{\partial x}\Big|_{x = \hat x_k^-}, & L_k &= \frac{\partial h}{\partial v}\Big|_{x = \hat x_k^-}.
\end{align}

The state $x$ and its error covariance $P$ can now be estimated iteratively as   
 \begin{align}
    \label{ekf_x_prior}
    \hat{x}_k^- &= f(\hat{x}_{k-1}^+, 0), \\
    \label{ekf_P_prior}
    P_k^- &= F_{k-1}P_{k-1}^+F_{k-1}^T + G_{k-1}QG_{k-1}^T, \\
    \label{ekf_gain}
    K_k &= P_k^-H_k^T(H_kP_k^-H_k^T + L_kRL_k^T)^{-1}, \\
    \label{ekf_x_post}
    \hat{x}_k^+ &= \hat{x}_k^- + K_k(z_k - h(\hat{x}_k^-, 0)), \\
    \label{ekf_P_post}
    P_k^+ &= (I - K_kH_k)P_k^-.
\end{align}

\subsection{Attitude Estimation Model}
The continuous-time model for attitude estimation using inertial sensors is
\begin{align}
    \label{cont_model}
    \dot{q} &= \frac{1}{2}q\otimes(\omega - \eta), \nonumber\\
    z&= \begin{bmatrix}
        a \\ m
    \end{bmatrix} = \begin{bmatrix}
        q^{-1} \otimes g \otimes q\\
        q^{-1} \otimes m_0 \otimes q
    \end{bmatrix} + \begin{bmatrix}
        \nu_a \\ \nu_m
    \end{bmatrix} \nonumber \\
    &=h(q) + \nu,
\end{align}
where $q$ is a unit-quaternion of the form $\begin{bmatrix}
    q_w & q_v^T
\end{bmatrix}^T$, with $q_w\in\mathbb{R}$ referred to as the `real part' of the quaternion and $q_v\in\mathbb{R}^3$ referred to as the `imaginary part' of the quaternion. The conjugate and inverse of a quaternion are defined similar to that of complex number: $\bar{q} \coloneq \begin{bmatrix}
    q_w & -q_v^T
\end{bmatrix}^T$ and $q^{-1} \coloneq \bar{q}||q||^{-2}$ respectively. Quaternion multiplication, denoted by $\otimes$, is called the Hamilton product. It can be encoded into matrix multiplication:
\begin{equation}
    p\otimes q = \Xi[p] q = \Omega[q] p,
\end{equation}
for quaternions $p$ and $q$, with $\Xi$ and $\Omega$ representing left and right multiplications respectively, defined as
\begin{align}
    \Xi[p] = \begin{bmatrix}
        p_w & -p_v^T \\ p_v & p_wI_3 + [p_v]_\times
    \end{bmatrix}, 
    \Omega[q] = \begin{bmatrix}
        q_w & -q_v^T \\ q_v & q_wI_3 - [q_v]_\times
    \end{bmatrix},
\end{align}
where $[v]_\times$, for $v=\begin{bmatrix}
    v_x & v_y & v_z
\end{bmatrix}$, is the skew-symmetric matrix
\begin{equation*}
    \begin{bmatrix}
    0 & -v_z & v_y \\
    v_z & 0 & -v_x \\
    -v_y & v_x &0
\end{bmatrix}.
\end{equation*}
Quaternion-vector multiplication is defined as
$q\otimes v \coloneq q \otimes \begin{bmatrix}
    0 & v^T
\end{bmatrix}^T$. 

The unit-norm constraint implies that $||q||^2 = q_w^2 + ||q_v||^2 = 1$. Only under this constraint does it represent a proper rotation. We can deduce that unit-quaternions are in bijection with points on the real 3-sphere $S^3$, which can be used to denote the set of all unit-quaternions: $S^3 = \left\{q\in\mathbb{H} : ||q||=1\right\}$, for $\mathbb{H} = \left\{\begin{bmatrix}
    q_w & q_v^T
\end{bmatrix}^T : q_w\in\mathbb{R},q_v\in\mathbb{R}^3\right\}$ being the set of all quaternions \cite{gma_ch9}. We can easily verify that for unit-quaternions, the conjugate and inverse are equivalent. For a primer to quaternion algebra and kinematics, readers may refer to Sections 1-4 in \cite{sola_quat}. 

$\omega \in \mathbb{R}^3$ is the angular velocity of the body, measured by the gyroscope, $a \in \mathbb{R}^3$ is proper acceleration in the body frame, measured by the accelerometer. $m \in \mathbb{R}^3$ is the magnetic field in the body frame, measured by the magnetometer. Under no acceleration and low magnetic interference, the measurements of the accelerometer and magnetometer are the earth's gravitational field $g$ and magnetic field $m_0$ respectively, rotated to the body frame. These measurements are corrupted by zero mean Gaussian noises, $\eta$ and $\nu$, defined by
\begin{align}
    \label{gyro_noise}
    \eta&\sim\mathcal{N}(0, \Sigma_\eta), \\
    \label{meas_noise}
    \nu_a&\sim\mathcal{N}(0, \Sigma_a), \quad \nu_m\sim\mathcal{N}(0, \Sigma_m), \nonumber\\
    \nu=\begin{bmatrix}
    \nu_a \\ \nu_m
\end{bmatrix}&\sim\mathcal{N}(0, \Sigma_\nu), \quad \Sigma_\nu = \begin{bmatrix}
        \Sigma_a & 0 \\
        0 & \Sigma_m
    \end{bmatrix}.
\end{align} 

Discretizing \eqref{cont_model} with time-step $\Delta t$, we get
\begin{align}
    \label{aekf_prop}
    q_{k+1} &= \left(I_4 + \frac{\Delta t}{2}\Omega[\omega_k]\right) q_k - \left(\frac{\Delta t}{2}\right)q_k\otimes\eta_k \nonumber\\
    &= \left(I_4 + \frac{\Delta t}{2}\Omega[\omega_k]\right) q_k - \left(\frac{\Delta t}{2} \Xi[q_k]\right)\eta_k,  \\
    \label{aekf_meas}
    z_k &= h(q_k) + \nu_k.
\end{align}

A first approach to attitude estimation would be to use equations $\eqref{aekf_prop}$ and $\eqref{aekf_meas}$ directly as the process and measurement models for the EKF, with the rotation quaternion as the state vector. While this approach is popular, dubbed the `additive extended Kalman filter' in literature, it poses several conceptual and practical problems.

The core conceptual flaw in the additive approach is the treatment of (unit-norm) quaternions as vectors, while a more rigorous approach would be to treat them as Lie groups, with closure under multiplication (not addition). Error too must be defined multiplicatively, not additively. Further, it is obvious from equation \eqref{aekf_prop} that $q_k$ is not Gaussian. In fact, it is the tangent space to the manifold $S^3$, not $S^3$ itself, that takes an (approximately) Gaussian form in the attitude estimation problem \cite{markley2004}.

The mathematics of Lie groups and random variables over Lie groups have been studied extensively, especially in the context of orientation and pose estimation \cite{bonnabel2008,falorsi2019,barfoot2014,chirikjian2012, barfoot2017}. Many Kalman filtering techniques on Lie groups and manifolds have also been developed, exploiting these mathematical properties \cite{bonnabel2010,bourmaud2015,bourmaud2013,hauberg2013,gui_liekf,hartley2020}. One such technique is the left-invariant extended Kalman filter (LI-EKF), which shall be the subject of this article.

\section{Robust LI-EKF}

\subsection{Left-Invariant Extended Kalman Filter}

Although Lie groups in the context of control theory have been studied since 1970s \cite{brockett73,Willsky75,Duncan77}, the invariant extended Kalman filter (IEKF) on Lie groups was introduced by Bonnabel \cite{bonnabel_likf} in 2007. The multiplicative extended Kalman filter (MEKF), a simplified and approximate version of the IEKF, was developed in NASA in 1969 for attitude estimation in multi-mission satellites \cite{paulson_spars,toda_spars,markley2004}. For a complete introduction to IEKF, readers may refer to \cite{barrau2016invariant,bonnabel_2018}.

As mentioned earlier, we define the error multiplicatively. For the estimated state $\hat{q}$ and the true state $q$, both being unit-quaternions, we can write the quaternion multiplicative error as
\begin{equation}
    \label{quat_err}
    \varepsilon = \hat{q}^{-1}\otimes q = \bar{\hat{q}}\otimes q.
\end{equation}
If $\hat{q}$ and $q$ are both unit-quaternions representing rotations, $\varepsilon$ is bound to be unit-norm and represent the `difference' between the two rotations. The nomenclature of the filter arises from the invariance of this error to left-multiplication: $\left(\Gamma\otimes\hat{q}\right)^{-1}\left(\Gamma\otimes q\right) = \hat{q}^{-1}\otimes q$ for any arbitrary quaternion $\Gamma$.

An exponential map $\mathbb{R}^3 \to S^3$ can be defined as
\begin{equation}
    \label{exp_def}
    Exp : \xi \mapsto \begin{bmatrix}
        \cos{||\xi||} \\
        \xi\frac{\sin{||\xi||}}{||\xi||}
    \end{bmatrix}.
\end{equation}

Substituting \eqref{quat_err} in \eqref{cont_model} and solving with $Exp(x/2) = \varepsilon$, we get
\begin{align}
    \label{alpha_def}
    \dot{x} &= -[\omega]_\times x - \eta,
\end{align}
where \eqref{alpha_def} is obtained from the approximation of small error: $Exp(\xi) \to \begin{bmatrix}
    1 & 0 & 0 & 0
\end{bmatrix}^T$ as $\xi \to \begin{bmatrix}
    0 & 0 & 0
\end{bmatrix}^T$.

Discretizing \eqref{alpha_def}, we get
\begin{align}
    \label{likf_prop}
    x_{k+1} &= \left(I_3 - \Delta t[\omega_k]_\times\right) x_k - \Delta t\eta_k.
\end{align}
The EKF can now be run with $x$ as the state vector. We still need to define the filter parameters $\{F, H, Q, R\}$.
The measurement model remains the same as in the additive case, given by equation \eqref{aekf_meas}. However, following EKF methodology, $h$ needs to be discretized with respect to $x$ (not $q$). Recall that $Exp(x/2) = \bar{\hat{q}}q$. The measurement sensitivity matrix $H$ can then be computed as
\begin{align}
    \label{H_likf}
    H_k = \frac{\partial}{\partial x}h(q)\Big|_{q = \hat q_k^-} 
    = \begin{bmatrix}
        [\hat{q}_k^{-^{-1}} \otimes g \otimes \hat{q}_k^-]_\times\\
        [\hat{q}_k^{-^{-1}} \otimes m_0 \otimes \hat{q}_k^-]_\times
    \end{bmatrix}.
\end{align}
From \eqref{likf_prop} and \eqref{aekf_meas}, we can write 
\begin{align}
    \label{F_likf}
    F_k &= I_3 - \Delta t[\omega_k]_\times,\\
    \label{noise_likf}
    Q &= \left(\Delta t\right)^{2}\Sigma_\eta,\quad 
    R = \Sigma_\nu.
\end{align}

With the parameters defined above and $x$ as the state vector, equations \eqref{ekf_x_prior}-\eqref{ekf_P_post} are iterated. Note that $x$ is not the orientation estimate, only a parameterization of the estimation error. The actual estimates are computed for each iteration as 
\begin{align}
    \label{likf_x_prior}
    \hat{q}_k^- &= \left(I_4 + \frac{\Delta t}{2}\Omega[\omega_{k-1}]\right)\hat{q}_{k-1}, \\
    \label{likf_x_post}
    \hat{q}_k^+ &= \hat{q}_k^-Exp\left(\frac{1}{2}K_k\left(z_k - h(\hat{q}_k^-)\right)\right).
\end{align}

\subsection{Noise Covariance Estimation}
\begin{table*}[t!]
\renewcommand{\arraystretch}{1.07}
\caption{Median of Root Mean Square Error Norm}
\label{table}
    \centering
\begin{tabular}{ |c|c|c|c|c|c|c|c|c| } 
 \hline
 & & \multicolumn{5}{|c}{With Adaptation} & \multicolumn{2}{|c|}{Without Adaptation} \\
 \cline{2-9}
  & $\Theta^0$& WL: 20 & WL: 40 & WL: 60 &WL: 80 &WL: 100 &with $\Theta_{true}$&with $\Theta^0$\\
 \hline\hline
 Median of & $\{Q^0, R^0\}$ &
$6.02\times10^{-3}$&
$5.79\times10^{-3}$&
$5.73\times10^{-3}$&
$5.69\times10^{-3}$&
$5.68\times10^{-3}$&
$5.60\times10^{-3}$&
$5.77\times10^{-3}$\\
 \cline{2-9}
 RMSE Norm& $\{{Q^0}', {R^0}'\}$&
$6.38\times10^{-3}$&
$5.95\times10^{-3}$&
$5.78\times10^{-3}$&
$5.82\times10^{-3}$&
$5.68\times10^{-3}$&
$5.60\times10^{-3}$&
$8.23\times10^{-3}$\\
 \hline
\end{tabular}
\end{table*}
As mentioned earlier, the sensor noise statistics might be unknown, hence $\Sigma_\eta$ and $\Sigma_\nu$ need to be estimated during filter operation. Expectation-maximization (EM) is a popular technique that has been applied to the joint state-parameter estimation problem in the context of Kalman filtering. Shumway and Stoffer \cite{shumway_em} first used the EM algorithm for noise covariance estimation in time-invariant linear Gaussian systems. Its convergence is described by Wu \cite{wu_em_convergence}. Bavdekar et al. \cite{bavdekar_em} reformulated the EM algorithm for non-linear systems using the extended Kalman filter and smoother. However, since in our case only the first term in equation \eqref{aekf_meas} is non-linear, we proceed with the estimator in \cite{shumway_em}, modified for the time-varying case and with a linearized $H$ matrix.

The LI-EKF is run with an initial parameter estimate $\Theta^0=\{Q^0,R^0\}$ over a window of length $n$. A Rauch–Tung–Striebel (RTS) smoother and a lag-one covariance smoother, which are given below, are then applied over the same window.

We initialize the RTS smoother as 
\begin{align}
    \label{rts_init}
    \hat x_n = \hat x_n^+,\quad
    P_n = P_n^+,
\end{align}
and iterate for $i = n-1,\dots,1,0$
\begin{align}
    \label{rts_iter}
    J_i &= P_i^+F_i(P_{i+1}^-)^{-1}, \nonumber\\
    P_i &= P_i^+ + J_i(P_{i+1} - P_{i+1}^-)J_i^T, \nonumber\\
    \hat x_i &= \hat x_i^+ + J_i(\hat x_{i+1} - \hat x_{i+1}^-).
\end{align}

Similarly, the lag-one covariance smoother is initialized as
\begin{align}
    \label{lag1_init}
    P_{n,n-1} &= (I-K_nH_n)F_{n-1}P_{n-1}^+,
\end{align}
and iterated for $i = n-1,\dots,1$
\begin{align}
    \label{lag1_iter}
    P_{i,i-1} &= P_i^+J_{i-1}^T + J_i(P_{i+1,i} - F_iP_i^+)J_{i-1}^T.
\end{align}

This is followed by the EM algorithm, consisting of two steps:
\begin{enumerate}
    \item Expectation: A log-likelihood function is formulated using the innovation $r_k = z_k - h(\hat{x}_k^-)$, expressed in terms of the present estimates of model parameters $\left(\Theta^j = \{Q^j,R^j\}\right)$ and the smoothed state and covariance estimates based on the present parameters:
    \begin{align}
        G(\Theta|\Theta^{j-1}) &= n\ln{|Q|} + n\ln{|R|} \nonumber \\&+tr\left\{Q^{-1}[S_{11} - S_{10} - S_{10}^T + S_{00}]\right\}\nonumber\\
        &+ tr\left\{R^{-1}\sum_{i=1}^n[(z_i - h(\hat{x}_i))(z_i - h(\hat{x}_i))^T\right. \nonumber\\
        &\qquad\left. + H_iP_iH_i^T]\right\}
    \end{align}
    where
\begin{align}
    \label{S}
        S_{11} &= \sum_{i=1}^n\left(\hat x\hat x_i^T + P_i\right),\nonumber \\
        S_{10} &= \sum_{i=1}^n\left(\hat x_i\hat x_{i-1}^T + P_{i,i-1}\right)F_{i-1}^T, \nonumber \\
        \text{and}\quad S_{00} &= \sum_{i=1}^nF_{i-1}\left(\hat x_{i-1}\hat x_{i-1}^T + P_{i-1}\right)F_{i-1}^T.
\end{align}
    \item Maximization: The log-likelihood function is minimized with respect to the parameters in the current iteration, yielding updated parameter estimates, given by
    \begin{align}
        \label{Q_est}
        Q^j &= \frac{1}{n}\sum_{i=1}^n\left[S_{11} - S_{10}S_{00}^{-1}S_{10}^T\right],\\
        \label{R_est}
        R^j &= \frac{1}{n}\sum_{i=1}^n\left[\left(z_i - h(\hat{x}_i)\right)\left(z_i - h(\hat{x}_i)\right)^T + H_iP_iH_i^T\right].
    \end{align}
\end{enumerate}
The Filter-Smoother-EM procedure is iterated over the same window until the convergence of the parameters is achieved, or until a maximum number of iterations is reached. The filter then proceeds with the estimated noise covariance matrices. 

\section{Simulation Results}

\begin{figure}
    \centering
    \includegraphics{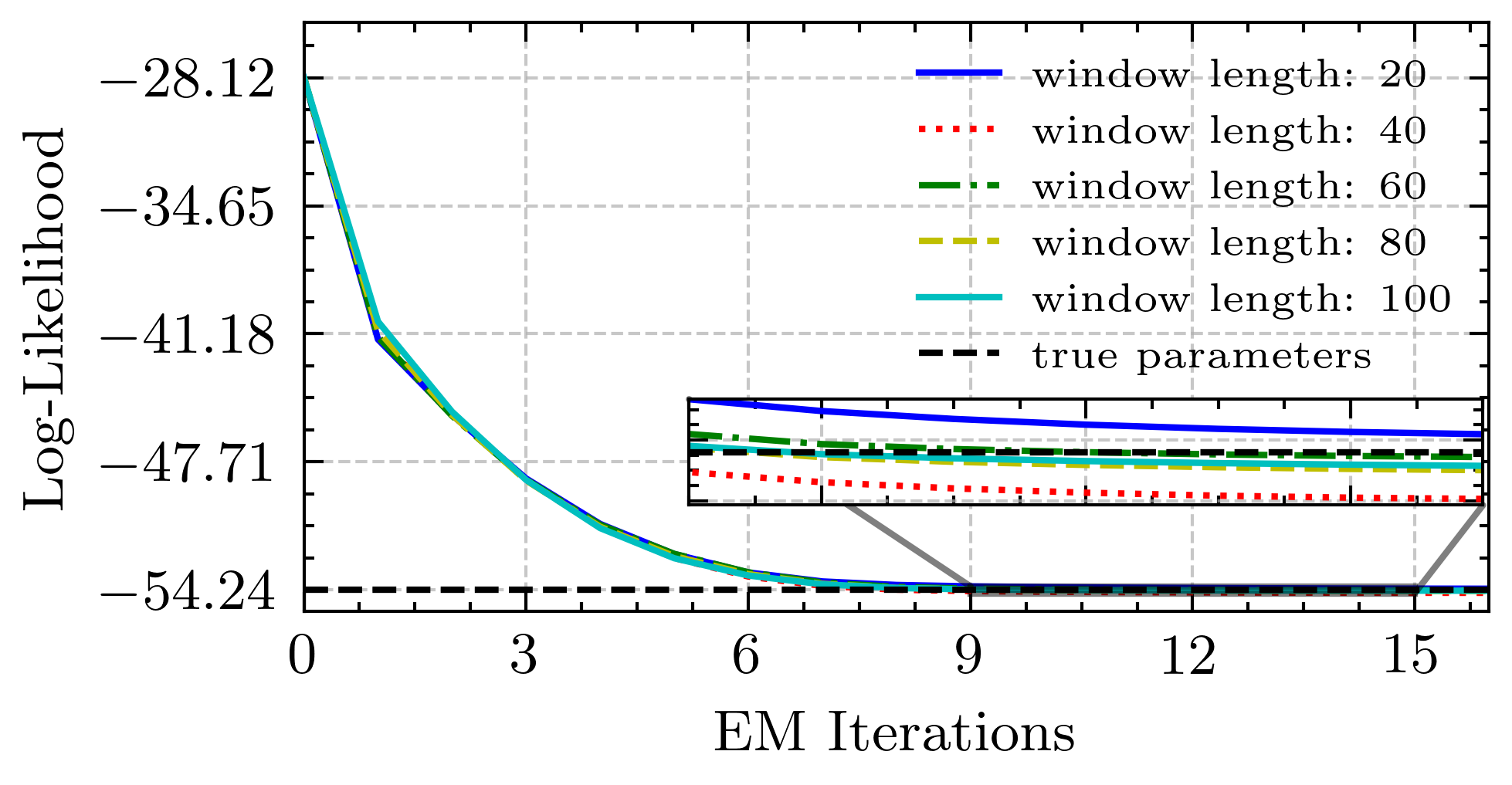}
    \caption{Convergence of the log-likelihood function.}
    \label{fig:cost_func}
\end{figure}

\begin{figure}
    \centering
    \includegraphics{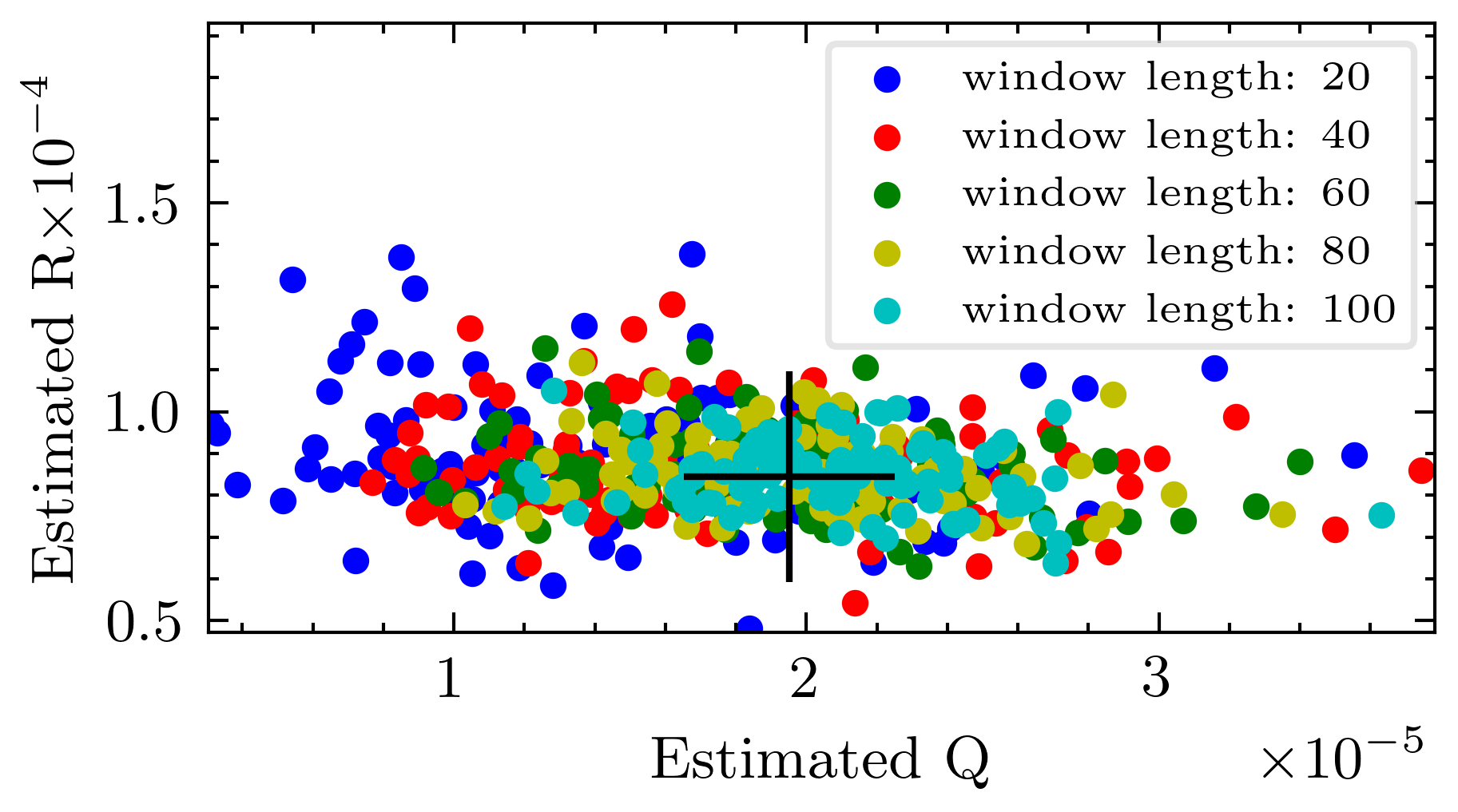}
    \caption{Scatter plot of Q and R estimates, {\huge+} denotes the true values.}
    \label{fig:qr_scatter}
\end{figure}

The LI-EKF along with the noise covariance estimation procedure is tested over 100 Monte Carlo runs with randomly generated noise added to a predefined trajectory (defined by $\omega,a$ and $m$). The noise terms $\eta$ and $\nu$ are generated with covariances $diag(\begin{bmatrix}
    0.75 & 1.5 & 1
\end{bmatrix})\times10^{-1}$ and $diag(\begin{bmatrix}
    1&2&3&3&3.5&6
\end{bmatrix})\times10^{-5}$ respectively. The timestep $\Delta t$ is set to $0.01$ s, giving $Q_{true} = (0.01)^2\Sigma_\eta=diag(\begin{bmatrix}
    0.75 & 1.5 & 1
\end{bmatrix})\times10^{-5}$ and $R_{true} =diag(\begin{bmatrix}
    1&2&3&3&3.5&6
\end{bmatrix})\times10^{-5}$.

For each Monte Carlo run, five LI-EKFs are iterated with different window lengths $n$ = 20, 40, 60, 80 and 100, for the noise covariance estimation procedure with initial parameters $\Theta^0=\left\{Q^0,R^0\right\} = \left\{400Q_{true}, 200R_{true}\right\}$. Two additional filters, without noise covariance estimation, are also run for comparison purposes, one with parameters $\Theta_{true}=\{Q_{true},R_{true}\}$ and the other with $\Theta^0$.

The minimization of the log-likelihood function (averaged over the window length) in a single run is shown in Figure \ref{fig:cost_func}. We observe that for all window lengths the functions converge close to the log-likelihood function for true parameters.

Figure \ref{fig:qr_scatter} shows the distribution of the Frobenius norm of the estimated $Q$ and $R$ for different window lengths. 

The median of the Euclidean norm of the RMSE across the 100 Monte Carlo runs is shown in Table \ref{table}, WL referring to the corresponding window lengths. Median RMSEs for filters with a different initial parameter estimate, $\{{Q^0}',{R^0}'\}=\{400Q_{true},0.2R_{true}\}$, are also shown. While the error for non-adaptive estimation varies dramatically with different parameter assumptions, the adaptive approach provides consistent error across initial estimates, remaining close to the error for the filter run with true parameters.

\section{Conclusion and Future Work}
In this article, we presented a robust quaternion LI-EKF framework for attitude estimation. The simulation results demonstrated the effectiveness of the proposed method, as the noise covariances converged to their true values across different initial estimates. Furthermore, the minimization of the log-likelihood function confirmed the accuracy of the noise estimation procedure.

Future work will focus on validating the algorithm in real-world dynamic environments and extending the framework to multi-sensor fusion problems.
\bibliographystyle{unsrt}
\bibliography{mybibfile}

\end{document}